# A biological approach to metalworking based on chitinous colloids and composites

*Ng Shiwei[1†], Ng Guan Zhi Benjamin[1†], Robert E. Simpson[2], Javier G. Fernandez[1]\**

[1] Engineering and Product Development, Singapore University of Technology and Design, 8 Somapah Road, 487372, Singapore

[2] School of Engineering, University of Birmingham, Edgbaston, Birmingham, B15 2TT, UK

† These authors contributed equally to this work

**E-mail:** javier.fernandez@sutd.edu.sg

**Abstract**

Biological systems evolve with minimum metabolic costs and use common components, and they represent guideposts toward a paradigm of manufacturing that is centered on minimum energy, local resources, and ecological integration. Here, a new method of metalworking that uses chitosan from the arthropod cuticle to aggregate colloidal suspensions of different metals into solid ultra-low-binder-content composites is demonstrated. These composites, which can contain more than 99.5% metal, simultaneously show bonding affinity for biological components and metallic characteristics, such as electrical conductivity. This approach stands in contrast with existing metalworking methods, taking place at ambient temperature and pressure, and being driven by water exchange. Furthermore, all the nonmetallic components involved are metabolized in large amounts in every ecosystem. Under these conditions, the composites' ability to be printed and cast into functional shapes with metallic characteristics is demonstrated. The affinity of chitometallic composites for other biological components also allows them to infuse metallic characteristics into other biomaterials. The findings and robust manufacturing examples go well beyond basic demonstrations and offer a generalizable new approach to metalworking. The potential for a paradigm shift toward biomaterials based on their unique characteristics and the principles of their manufacturing methods is highlighted.

**1. Introduction**

The field of bioinspired materials represents our attempt to learn from nature's design of biological composites, which result from millennia of evolution. Understanding how biological systems integrate complex hierarchical designs on multiple length scales enables the replication of the same principles and novel functionalities using artificial materials[1, 2]. However, beyond the adaptation of specific strategies at the product level, we are still in the nascent phase of applying the essential lessons taught by nature's use of local components, favor of activities with minimal metabolic energy, and seamless integration within Earth's ecological cycles[3], which could ultimately lead to a biological transformation of manufacturing systems[4]. Unlike the modern manufacturing paradigm, which depends upon complex global supply chains and seemingly consequence-blind resource exploitation, natural systems have evolved to produce outstanding functional structures from common components while minimizing the metabolic cost[5]. Manufacturing with metals is a paradigmatic example of the different approaches to artificial and biological production.

The existing approach to manufacturing metals is based on melting and shaping processes requiring substantial energy sources to reach high temperatures and pressures. These processes are not unique to metalworking, rather they are a signature of our path of industrialization, which is characterized by an ever-increasing dependence on highly energy-demanding processes[6]. Our use of metals within a paradigm of significant energy sources contrasts with the rare, but real, use of metals in biological structures. Found and studied mainly in the chitinous cuticles of arthropods[7], these metals (mostly Zn, Mn, and Fe) are incorporated at ambient conditions into critical areas, such as the tips of the claws and fangs or the shells of eggs, to exploit their mechanical properties[8-10]. Metals, however, are not generally incorporated into arthropod cuticles at the first stage of molting[11]. Initially, the organic phase is secreted into a water-based environment, producing a soft hydrogel that transitions to a stiff shell through tanning and controlled dehydration. The anisotropic shrinking and associated strong intermolecular forces generated by the removal of water are essential aspects of the cuticle's consolidation and extraordinary mechanical properties[12, 13]. Indeed, they compel the rearrangement of chitinous molecules and fibers to strengthen their intermolecular bonds[13, 14]. It is in this late stage of development when metals are incorporated from the environment[15]. While this metal embedded in the cuticle is generally associated with proteins, chitinous polymers also show a strong affinity for metals, which are retained after their extraction. In fact, one of the most extensive uses of chitinous waste, such as that from the fishing industry, is in flocculating heavy metals in water remediation systems[16].

In this paper, we explain how this affinity of chitinous biomolecules for metals goes far beyond the removal of metallic elements from water. We show the aggregation of metal particles from a colloidal suspension into a solid composite without using external pressure, thus creating an agglomeration that is hundreds of times the weight of the original chitin. This process enables conventional manufacturing techniques, such as casting, coating, and 3D printing, to be adapted to a new context of biological conditions (i.e., metabolizable chemicals under ambient temperature and pressure) in a unique example of a large-scale, ecologically integrated, and generalizable approach to metalworking.

The proposed approach, which is heavily influenced by the processes that allow the use of metals in the cuticle, prioritizes the adaptation of biological tools and strategies to meet the requirements of an industrial production system that is integrated with biological systems and cycles. It does not aim to precisely replicate the exact biological processes that enable the use of metals in chitinous animals. Therefore, there are several significant differences between the described results and the incorporation of metals in the cuticle: (i) Metals in the chitinous cuticle are commonly incorporated through binding to the histidine residues of proteins, possibly via chelation at the imidazole functional group[15, 17]. Here we use the metal affinity of chitosan as the main bridge of the organic-inorganic links[18], favoring the use of a ubiquitous molecule to achieve industrially relevant production scales[19]. (ii) In the arthropod cuticle, large amounts of metals are incorporated at the molecular level at concentrations that can reach 25% of the cuticle weight by the end of the animal's life[20]. Unbounded by the limits of metal bioavailability, we explore concentrations that are orders of magnitude larger than those in the cuticle and the use of metallic particles, thus enabling the formation of organometallic composites that have the fundamental functional properties of metals rather than those of a modified organic material[21-23]. (iii) We use some of the principles behind chitinous cuticle stiffening, which is achieved at ambient temperature and pressure, to consolidate the composite from a colloidal form to a compacted porous metal through water exchange at biological conditions, embodying an approach that can be incorporated into other bioinspired manufacturing technologies[24]. However, in the arthropod cuticle, metals are incorporated after or during the consolidation of the organic matrix, not before.

## 2. Results

### 2.1. From colloids to solids

The process of producing chitometallic composites is illustrated in **Figure 1a.** Briefly, chitosan, which was extracted from shrimp, was dissolved at a 3% concentration in a weak acetic acid solution at the anaerobic fermentation limit[25] (1% v/v, table vinegar ranges between 4 and 8%). A key factor in this dissolution is that, despite its considerable molecular weight (150–200 kDa), chitosan can be dispersed by the presence of easily protonated amine groups, enabling the introduction of repulsive inter- and intramolecular forces. Notwithstanding the low concentration of chitosan, the resulting solution behaves as a liquid crystal[26], simultaneously able to flow as a liquid and to interact intermolecularly at long range like a solid.

In this state, we added metallic particles of tin, copper, and stainless steel in varying ratios (from 50 to 250 times the weight of dry chitosan), forming colloidal suspensions. Copper and tin were selected as examples because they are key metals in electrical engineering and because copper is a transition metal while tin is a non-transition metal, each thus having a different chelation mechanism with chitosan[27]. Stainless steel was selected because of its known inert nature and its lack of interaction with chitosan, requiring a surface modification to show affinity[28]. As the water evaporated, the rearrangement of the chitosan chains, the new intermolecular direct bonds, and overall shrinking due to the lost volume resulted in strong forces (**Figure 1b**)[12, 29], internally compacting the metallic particles into a granular solid. For all three metals, the result was a compact but porous aggregate of metallic particles (**Figures 1c, d**). An illustrative example of the forces exerted by chitosan during dehydration—which, in the arthropod cuticle, provides stiffness and uniaxial strength through strain hardening[30]—can be found in our previous work in which these forces were employed to move large architectural constructions in response to ambient humidity changes[31].

The minimum amount of chitosan required to consolidate the particles into a solid ranged from 0.4% to 1.0% of the metal's weight and was largely independent of the metal chosen, with particle size being the determining factor. This suggests that the formation of the solid is likely to be a predominantly physical process (i.e., rather than chemical) governed by the ability of chitosan to bridge and reduce the interstitial space between particles. The concentrations of chitosan necessary to bind metallic particles are lower by an order of magnitude than the binder concentrations in previously reported ultra-low-binder-content (UBC) composites[32]. However, unlike UBC composites, which are made using synthetic polymers and high external pressures, the chitometallic composites reported here achieved these results without external forces and under ambient conditions, highlighting the unparalleled efficiency of this biological approach to bind particles.

## 2.2. Casting and extending metallic properties to other biocomposites

Casting is an essential metal manufacturing process that, until now, requires high temperatures. However, the unique ability of the chitometallic colloids presented here to "self-aggregate" into solids from a flowing state allows the casting of metal composites at ambient conditions. To illustrate this, we printed 3D negative molds of different geometrical shapes and filled each with a colloidal suspension. We then allowed the vitrification of chitosan and thus allowed consolidation of the composite in the mold, resulting in a metallic replica (**Figure 2a**). Casting a colloidal suspension into a solid requires a balance of the properties of both states. In the colloidal form, the initial concentration of chitosan in the solution determines rheology and manufacturability, where an excess of dissolved chitin makes it easier for the colloid to conform to a shape but harder to retain that shape after shrinking. Likewise, the chitosan:metal ratio influences the characteristics of the solid form (**Figure 2b**), as low amounts of chitosan result in loose aggregates, while an excess of chitosan results in the metallic electrical, thermal, and abrasion resistance properties of the filler being overshadowed by the properties of the chitosan.

All three tested metals produced complex shapes (**Figure 2c**), confirming the general ability of chitosan to bind metals[33]. However, although the three metals were cast under the same conditions, copper reacted with the diluted acetic acid in the colloid to form copper (II) acetate. Phase separation was therefore induced and noticeably significant during the drying process, resulting in lower detail and low mechanical strength. This effect was partially overcome by neutralizing the excess acetic acid before introducing the metal. Therefore, further applications of the technology described here should consider the targeted metals' interactions, not only with the biopolymer, but also with the dispersing medium.

Compared to continuous (i.e., melted) pieces of metal, the porous chitometallic composites have a very poor standalone mechanical strength. However, they acquired enhanced resistance to thermal degradation (**Figure S1**) and abrasion from their metallic elements. The latter enables the pieces to be polished and thus acquire a characteristic metallic shine and, more importantly, surface continuity and metallic electrical conductivity (**Video S1**). Interestingly, despite the natural hydrophilicity of chitosan, the chitometallic composites show unanticipated moisture stability, with no observable mass loss or shape change after several days immersed in water (**Figure S2**). Furthermore, despite the low amount of biopolymer, the chitometallic composites retain chitosan's compatibility with biological composites, which, in addition to the ability to be formed at ambient temperatures, enables the localized incorporation of chitometallic properties into other biomaterials. We used this property to provide metallic characteristics to inexpensive, sustainable, large-scale constructions of cellulose-based fungal-like adhesive materials (FLAM)[34].

**Figure 2d** shows the process of incorporating the properties of tin onto the surface of two solid cellulosic objects: a replica of a honeybee's (*Apis mellifera*) head and the Merlion mascot of Singapore (**Figure 2e**). Because of the natural affinity of chitosan's amine groups to biological materials, FLAM objects can be dipped or "painted" in a 1:250 chitinous colloidal suspension of tin. The colloid self-consolidates (i.e., vitrifies) in a conforming metallic layer that can then be polished or reworked (**Figure 2f**, **Video S2**). Notably, since FLAMs are based on the same bioinspired manufacturing principles and common biomolecules as the metallic colloid, the entire manufacturing process—including the formation of the cellulosic insect and Merlion—was done at room temperature, simply using water to drive the assembly.

The chitinous metallic coating of cellulosic constructs goes beyond merely aesthetic or electrical modifications. In the arthropod cuticle, metallic deposits are incorporated at sharp tips to increase their durability[10, 35]. Inspired by this localized use of metal, we produced a sharpened FLAM tip and dip-coated it with tin. This simple process resulted in a modified sharp tip that had twice the yield stress of an uncoated tip and that required more than five times the energy to suffer irreversible damage (i.e., plastic deformation) (**Figure 2g**). It is worth noting that the affinity of chitosan for biological construction is not limited to cellulose. Most notably, the straightforward use of the principles described here to provide electrical properties to other biomaterials that have a demonstrated affinity with chitosan, such as internal organs, skin[36], and cotton[37], suggests their potential inclusion in smart textiles and medical devices.

### 2.3. 3D printing at ambient temperature and heat treatment

So far, we have demonstrated that chitometallic composites produced under ambient conditions allow electrical conductivity along the surfaces of solid chitometallic or cellulosic objects, and these can easily be used to create conductive tracks and patterns (**Video S3**). However, these metallic properties were only apparent after polishing, suggesting that the porous interior lacks continuity. We now show how chitometallic structures can become internally electrically conducting.

We achieved internal continuity, which would be needed for applications that require bulk electrical properties, such as batteries and electrodes, by exposing the chitometallic composites to increased temperatures for a short time. After heat treatment at temperatures typically lower than 300 °C, we observed a drastic reduction in electrical resistance in both tin and copper composites, while stainless steel composites were unaffected (**Figure S3**). This change in conductivity, which was ten orders of magnitude for tin and six orders of magnitude for copper, was initially hypothesized to result from the carbonization of chitosan, reduction of a metal oxide, or sintering. However, a comparison between thermal analyses of all the involved components and the temperature at which the composites attain conductivity suggested otherwise (**Figures 3a, b**). The existence of a critical threshold for inducing a permanent change in the bulk material at a temperature that has no apparent relevance to any of the involved components suggests further internal compaction and the achievement of continuity by percolation. This effect was particularly evident in copper composites, where bulk electrical continuity can be attained at temperatures as low as 172 °C, a fraction of the melting point of copper, and a temperature before chitosan, which is an electrical insulator, is thermally degraded (**Figure S3**). The exception of stainless steel, which can be aggregated during the vitrification of chitosan but cannot be further packed by the same treatment as copper and tin, suggests that the affinity of the metal for the chitinous binder might be a fundamental aspect of this process. The percolation hypothesis agrees with the SEM-EDS analysis and elemental analysis, which revealed low traces of carbon and distinct metallic particles with closer packing than before heat treatment (**Figure 3c, Table S1**). Similar results were obtained by mercury intrusion porosimetry, where, despite the loss of chitosan at high temperatures, we observed a significant reduction in interstitial porosity and an increase in bulk density for the heat-treated samples, confirming the reduction of interstitial space between metallic particles after heat treatment (**Figure S4**). While the cumulative evidence points to a percolation process, the fact that the transition was more obvious in the metal with the lowest melting point (i.e., tin) suggests that there could be a combination of effects. Further experiments, which are beyond the scope of the current study, will be needed to confirm these observations and to fully understand the phenomenological aspects of the process through which chitometallic composites acquire bulk electrical properties.

Safely manipulating and drying the material in ambient conditions and then applying a heat treatment as an optional post-process allows metal elements to be shaped without high heat. This opens a more accessible and flexible approach to metalworking, making it easier to produce shapes and, for example, to bioprint functional electrical components under ambient conditions.

Bulk 3D printing with the tin–chitosan colloid was found to be particularly straightforward because of its shear-thinning characteristics, which facilitate extrusion and help it to retain its shape thereafter. We pneumatically extruded the colloid using a standard 150 mL plastic syringe, actuated by computer-controlled stepper motors assembled on a cartesian system to control the position (**Figure S5**). Among the several objects that were printed with this system in order to demonstrate its ability to produce functional electrical components was a bulb socket, shown in **Figure 3d**. The process consisted of two well-defined stages: bioprinting and heat treatment, with the latter performed by raising the temperature of the printing surface to 320 °C for 10 min. The success of the process was demonstrated visually by lighting a bulb. A more detailed, uncut version of the process is shown in **Video S4**. Although we conducted the whole process in an open atmosphere, the heating of some metals might lead to the formation and reduction of oxidated states, which would require, in those cases, the use of controlled (e.g., neutral) atmospheres.

## 3. Discussion and outlook

In this paper, we have described the concepts and technological applications of an approach to metalworking that is based on water exchange and biological components, rather than high temperatures. The main advancements achieved here are: (1) a non-specialized and low-energy manufacturing process that can be scaled to industrial needs; (2) capitalizing on a ubiquitous resource and demonstrating its untapped utility; and ultimately (3) a realized technological step within a broader biological approach aiming towards sustainable development[24].

Traditionally, working with metals has been characterized by high-temperature processes that have been defined by the melting temperatures of the metals in use. In contrast, the approach presented here is characterized by the frugality of the processes in biological systems[5], and it has been demonstrated on a large scale, using low-quality (inconsistent shapes and sizes) and low-cost metallic particles, at ambient temperature and pressure, and —for the optional thermal treatments—under atmospheric conditions. Decoupling the thermal treatment from manufacturing enables a new and more accessible approach to working with metals. We have demonstrated how this enables models of metal manufacturing that are safe for users and the environment without specialized sintering equipment, strong solvents, or industrial facilities. Moreover, these low-temperature processes pose a solution to form electrical connections on heat-sensitive substrates, such as flexible electronics[38]. This approach also has the potential to integrate easily with other biological components and systems operating under ambient conditions.

The bridge between the biological and metal domains—chitosan—is a common biopolymer that can be derived from multiple sources—including urban waste[39]—within a circular, sustainable economy and can be safely handled without toxic chemicals. We showed that chitosan's extraordinary ability to integrate metals enables the production of functional organic–inorganic composites under ambient conditions from colloidal suspensions of metal powders.

The differences between the proposed approach and traditional metalworking are significant at a fundamental level, defining a technology that does not replace the old, but enables a new complementary production paradigm based on the biological transformation of industrial manufacturing[40]. The proposed technology is separate from typical applications of traditional metalworking, like the fabrication of standalone load-bearing structures, but overlaps with other uses, such as producing durable electrical components. It also enables new fields, such as integration with biological systems and sustainable modes of production. While modern society is built on high-performance technical materials, alloys, and composites, an enduring society would need to be built on materials and technologies that also seek to preserve. As this study demonstrates, nature has a library of strategies and materials that, if wisely employed, could bring us to a new technological age[24].

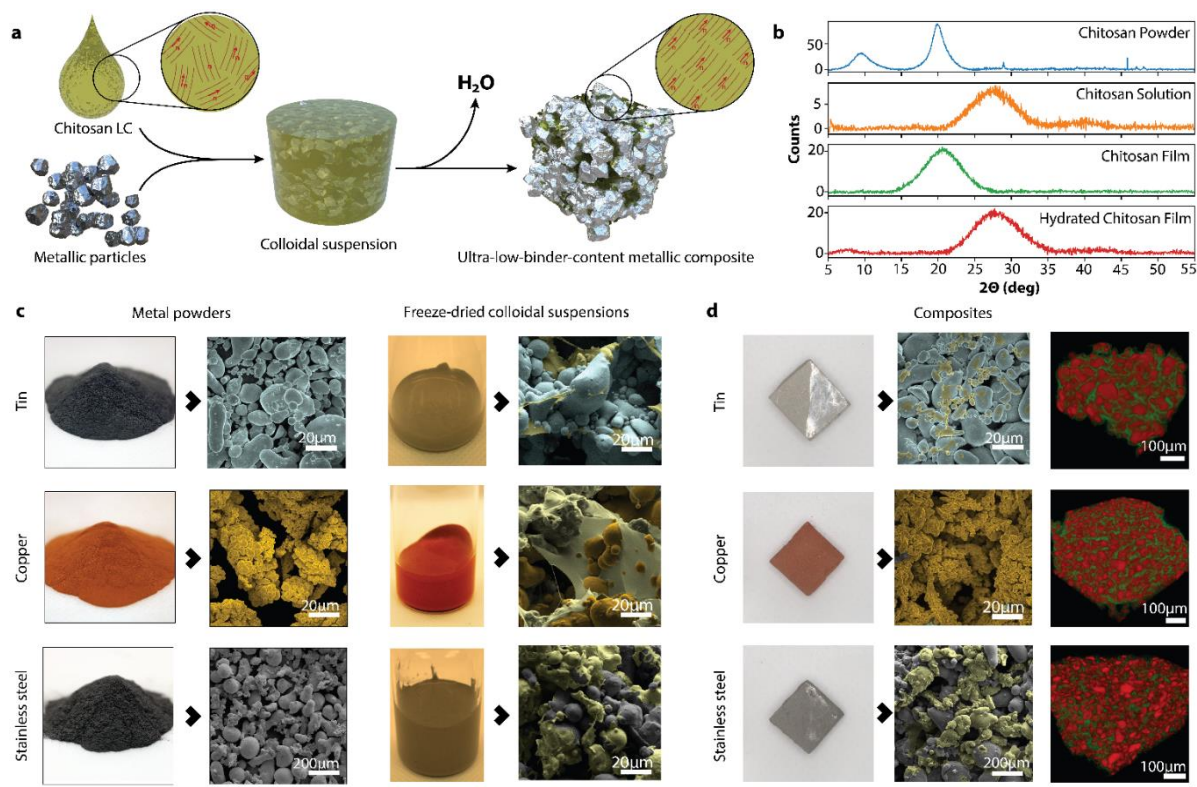

**Fig.1| Chitinous metallic colloids and ultra-low-binder-content composites. a,** Diagram of the formation of chitometallic colloidal suspensions and their transformation into ultra-low-binder-content composites by internal compaction. A liquid crystal solution of chitosan is used as the continuous phase, to which metallic particles are added as the dispersed phase. The resulting colloidal suspension, which can flow into and retain shapes, transitions to a solid as the water evaporates under ambient conditions. During this process, the vitrification of the relatively low amount of chitin (<0.4% w/w) brings all the metallic particles together into a porous aggregate that exhibits metallic characteristics. **b,** X-ray diffraction spectra of chitosan. The top row shows the spectrum of the raw chitosan extracted from shrimp shells, with the typical L-2 chitosan polymorph. The next three rows present the spectra of the liquid crystal chitosan and the evaporated film. Note the change in crystallite size with the water content in the bottom two rows. **c,** The two left columns are images of the three types of metallic particles employed here, showing the diversity of sizes and aggregation states. The two right columns are images of the colloidal suspensions and their colored scanning electron images after freeze-drying. The right-most column shows the initial intermingled state between particles and biopolymer and the existing connections between the chitinous molecules in suspension. **d,** Images of the solid chitometallic composites. The left column shows photographs of 2×2cm square composite samples of different metals. The central column shows scanning electron micrographs in which the chitosan has been colored. The right column shows reconstructed volumes of the composites recorded by x-ray microtomography, illustrating the high porosity of these chitometallic aggregates.

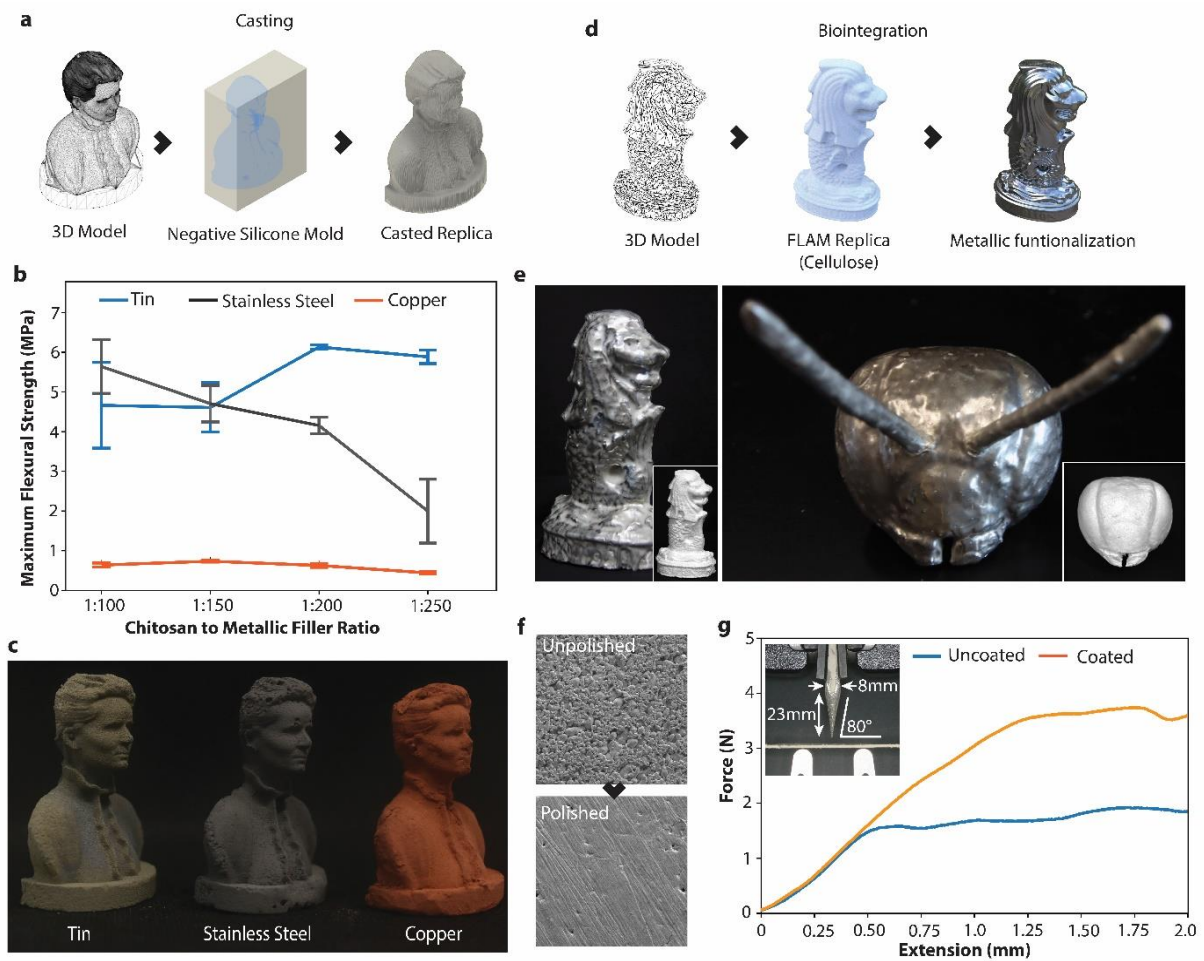

**Fig.2| Manufacturing of chitinous and chitinocellulosic metallic objects at standard temperature and pressure. a,** Diagram of a casting process using a chitometallic colloid. In the example, the model was printed in nylon using a commercial selective laser sintering (SLS) 3D printer for rapid prototypes of high-quality finish. A flexible silicone mold in the shape of the object's negative was then produced, which was used to cast the colloidal suspension. **b,** Flexural strength of the metallic aggregates with respect to the chitosan-to-metal ratio. All samples were cast under the same conditions. In the case of copper, the poor mechanical strength resulted from premature precipitation of the metal due to the slightly acidic chitinous continuous phase. **c,** Results of casting models of 5 cm tall Marie Curie in tin, stainless steel, and copper. **d,** Process for using the chitometallic colloids to incorporate metallic properties into other biocomposites, specifically the coating of cellulosic objects. **e,** Example of two large objects, a 13 cm tall Merlion, and a 7cm tall (without antennae) bee head, produced in fungal-like adhesive material (FLAM) and coated with tin under ambient conditions. Insets show the cellulosic objects before coating. The fabrication process is depicted in detail in Video S2. **f,** Scanning electron images of the surface of a chitometallic object before and after polishing. **g,** Representative measurement of the mechanical performance of a dip-coated FLAM tip. In the chitinous arthropod cuticle, metallic composites are used to enhance durability rather than to form structures. The coated tip in the image doubles its elastic region (linear slope) when crushed against a solid surface, requiring more than five times the energy (area under the curve in the elastic region) to undergo permanent deformation.

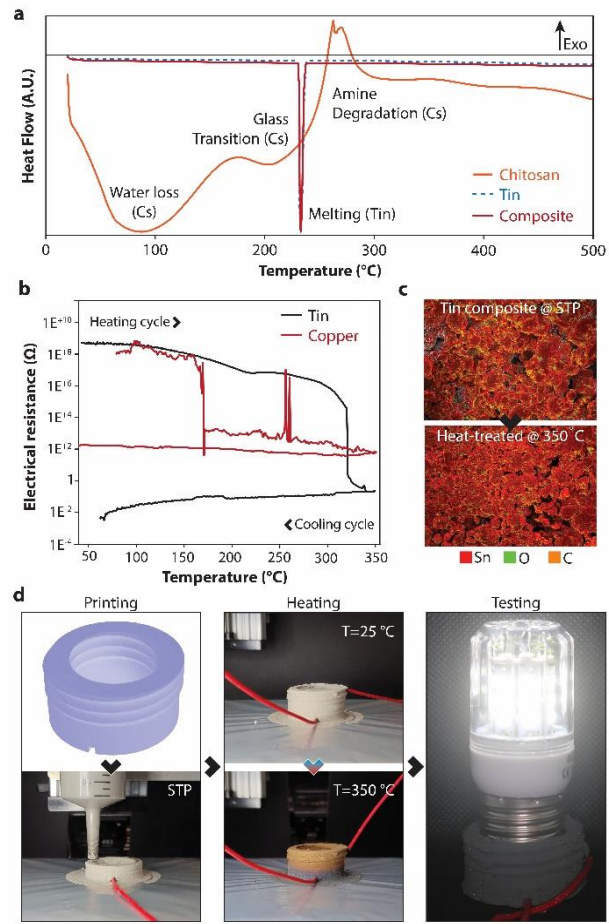

**Fig.3| Thermal and electrical properties. a,** Heat flow of the chitosan film, tin, and chitosan–tin composite, as measured by differential scanning calorimetry. Due to only small amounts of chitin in the composite, there is no observable difference in the intrinsic thermal characteristics of the composite and the constituent metal. **b**, Effect of thermally treating tin- and copper-based chitometallic objects. The composites transition to being electrical conductors at moderate temperatures, showing a permanent decrease in their electrical resistance of several orders of magnitude. This permanent increase in electrical conductivity occurred in all the tested metals at temperatures that have no relevance to the metal in question. Instead, the transition temperature appears dependent on aspects of the packing of the composite, suggesting the existence of an internal consolidation mechanism that is driven by the chitinous phase and goes beyond the percolating threshold required to achieve volumetric electrical continuity. For further information and a discussion of the process in tin and other metals, refer to Figure S3 and Table S2. **c,** Physical appearance and composition of a chitometallic composite based on tin, before and after thermal treatment, as measured by energy dispersive spectroscopy. **d,** Bioprinting of a bulb socket (24 mm tall, 25 mm I.D., 34 mm O.D.) at room temperature using a chitometallic colloid, followed by its thermal treatment in an open atmosphere to achieve electrical functionality. For further details on this fabrication, refer to Figure S5 and Video S4. (STP refers to standard temperature and pressure)

## Supporting Information

Supporting Information including materials and methods, five additional figures, two tables and four videos is available from the Wiley Online Library.

## Acknowledgements

We would like to thank Dr. Ludger Paul Stubbs and Mr. Ng Qi Hua for their assistance in performing x-ray microtomography measurements at the Institute of Sustainability for Chemicals, Energy, and Environment. We would like to thank Ms. Jing Ning and Ms. Niu Jiabin for their assistance with the x-ray diffraction, Ms Wei Ling for the elemental analysis and Dr Hong Lu for the mercury intrusion porosimetry. We would like to thank Hank Atar and Joao Burini for producing the 3D models of Marie Curie and the bee's head, and Chitinous Pte. Ltd. for producing the FLAM replicas.

## Contributions

NGZB and JGF conceived the idea of creating chitometallic composites. NGZB performed the proof of concept of composites and thermal treatments. NS performed the physical characterizations and developed the casting. NGZB developed the printing device and methods. JGF performed the coatings. RES developed the electrical measurements and tools. JGF and NS wrote the first draft of the article. All authors contributed to writing the manuscript.

## Conflicts of interest

The technology described in this document and its applications are part of a patent application in which NGBZ, NS, and JGF are inventors.

## References

1. Aizenberg, J. and P. Fratzl, *Biological and Biomimetic Materials.* Advanced Materials, 2009. **21**(4): p. 387-388.

2. Aizenberg, J. and P. Fratzl, *New Materials through Bioinspiration and Nanoscience.* Advanced Functional Materials, 2013. **23**(36): p. 4398-4399.
3. Vincent, J.F.V., *Biomimetic materials.* Journal of Materials Research, 2008. **23**(12): p. 3140-3147.
4. Miehe, R., et al., *The biological transformation of industrial manufacturing – Technologies, status and scenarios for a sustainable future of the German manufacturing industry.* Journal of Manufacturing Systems, 2020. **54**: p. 50-61.
5. Vincent, J.F.V., *Survival of the cheapest.* Materials Today, 2002. **5**(12): p. 28-41.
6. Jordan, P.G., *8 - Global Markets*, in *Solar Energy Markets*, P.G. Jordan, Editor. 2014, Academic Press: Boston. p. 127-133.
7. Vincent, J.F.V., *Arthropod cuticle: a natural composite shell system.* Composites Part A: Applied Science and Manufacturing, 2002. **33**(10): p. 1311-1315.
8. Lehnert, M.S., et al., *An augmented wood-penetrating structure: Cicada ovipositors enhanced with metals and other inorganic elements.* Scientific Reports, 2019. **9**(1): p. 19731.
9. Schofield, R. and H. Lefevre, *Short Communication: High Concentrations of Zinc in the Fangs and Manganese in the Teeth of Spiders.* Journal of Experimental Biology, 1989. **144**(1): p. 577-581.
10. Politi, Y., et al., *A Spider's Fang: How to Design an Injection Needle Using Chitin-Based Composite Material.* Advanced Functional Materials, 2012. **22**(12): p. 2519-2528.
11. Schofield, R.M.S., M.H. Nesson, and K.A. Richardson, *Tooth hardness increases with zinc-content in mandibles of young adult leaf-cutter ants.* Naturwissenschaften, 2002. **89**(12): p. 579-583.
12. Rukmanikrishnan, B., K.J. Tracy, and J.G. Fernandez, *Secondary Reorientation and Hygroscopic Forces in Chitinous Biopolymers and Their Use for Passive and Biochemical Actuation.* Advanced Materials Technologies, 2023. **8**(18): p. 2300639.
13. Vincent, J.F.V. and J.E. Hillerton, *The tanning of insect cuticle—A critical review and a revised mechanism.* Journal of Insect Physiology, 1979. **25**(8): p. 653-658.
14. Glaser, A.E. and J.F.V. Vincent, *The autonomous inflation of insect wings.* Journal of Insect Physiology, 1979. **25**(4): p. 315-318.
15. Politi, Y., et al., *Nano-channels in the spider fang for the transport of Zn ions to cross-link His-rich proteins pre-deposited in the cuticle matrix.* Arthropod Structure & Development, 2017. **46**(1): p. 30-38.
16. Skaugrud, E.O.O., *Metal recovery using chitosan.* Journal of Chemical Technology & Biotechnology, 1990. **49**(4): p. 395–404.
17. Khare, E., et al., *Molecular understanding of $Ni^{2+}$-nitrogen family metal-coordinated hydrogel relaxation times using free energy landscapes.* Proceedings of the National Academy of Sciences, 2023. **120**(4): p. e2213160120.
18. Nie, J., Z. Wang, and Q. Hu, *Chitosan Hydrogel Structure Modulated by Metal Ions.* Scientific Reports, 2016. **6**(1): p. 36005.
19. Ng, S., B. Song, and J.G. Fernandez, *Environmental attributes of fungal-like adhesive materials and future directions for bioinspired manufacturing.* Journal of Cleaner Production, 2021. **282**: p. 125335.
20. Schofield, R.M.S., et al., *Zinc is incorporated into cuticular "tools" after ecdysis: The time course of the zinc distribution in "tools" and whole bodies of an ant and a scorpion.* Journal of Insect Physiology, 2003. **49**(1): p. 31-44.
21. Khare, E., N. Holten-Andersen, and M.J. Buehler, *Transition-metal coordinate bonds for bioinspired macromolecules with tunable mechanical properties.* Nature Reviews Materials, 2021. **6**(5): p. 421-436.

22. Kim, S., et al., *In situ mechanical reinforcement of polymer hydrogels via metal-coordinated crosslink mineralization.* Nature Communications, 2021. **12**(1): p. 667.
23. Chou, C.-C., et al., *Ion Effect and Metal-Coordinated Cross-Linking for Multiscale Design of Nereis Jaw Inspired Mechanomutable Materials.* ACS Nano, 2017. **11**(2): p. 1858-1868.
24. Fernandez, J.G. and S. Dritsas, *The Biomaterial Age: The Transition Toward a More Sustainable Society will Be Determined by Advances in Controlling Biological Processes.* Matter, 2020. **2**(6): p. 1352-1355.
25. Park, Y.S., et al., *Production of a high concentration acetic acid by Acetobacter aceti using a repeated fed-batch culture with cell recycling.* Applied Microbiology and Biotechnology, 1991. **35**(2): p. 149-153.
26. Terbojevich, M., et al., *Chitosan: chain rigidity and mesophase formation.* Carbohydrate Research, 1991. **209**: p. 251-260.
27. Guibal, E., *Interactions of metal ions with chitosan-based sorbents: a review.* Separation and Purification Technology, 2004. **38**(1): p. 43-74.
28. Yuan, S., et al., *Enhancing antibacterial activity of surface-grafted chitosan with immobilized lysozyme on bioinspired stainless steel substrates.* Colloids and Surfaces B: Biointerfaces, 2013. **106**: p. 11-21.
29. Agarwal, U.P., et al., *Effect of sample moisture content on XRD-estimated cellulose crystallinity index and crystallite size.* Cellulose, 2017. **24**(5): p. 1971-1984.
30. Vincent, J.F. and U.G. Wegst, *Design and mechanical properties of insect cuticle.* Arthropod structure & development, 2004. **33**(3): p. 187-199.
31. Gupta, S.S., et al., *Prototyping of chitosan-based shape-changing structures.* 2019.
32. Oh, K., et al., *Compaction self-assembly of ultralow-binder-content particulate composites.* Composites Part B: Engineering, 2019. **175**: p. 107144.
33. Krajewska, B., *Diffusion of metal ions through gel chitosan membranes.* Reactive and Functional Polymers, 2001. **47**(1): p. 37-47.
34. Sanandiya, N.D., et al., *Large-scale additive manufacturing with bioinspired cellulosic materials.* Scientific Reports, 2018. **8**(1): p. 8642.
35. Schofield, R.M.S., et al., *The homogenous alternative to biomineralization: Zn- and Mn-rich materials enable sharp organismal "tools" that reduce force requirements.* Scientific Reports, 2021. **11**(1): p. 17481.
36. Fernandez, J.G., et al., *Direct Bonding of Chitosan Biomaterials to Tissues Using Transglutaminase for Surgical Repair or Device Implantation.* Tissue Engineering Part A, 2016. **23**(3-4): p. 135-142.
37. Chung, Y.-S., K.-K. Lee, and J.-W. Kim, *Durable Press and Antimicrobial Finishing of Cotton Fabrics with a Citric Acid and Chitosan Treatment.* Textile Research Journal, 1998. **68**(10): p. 772-775.
38. Espera, A.H., et al., *3D-printing and advanced manufacturing for electronics.* Progress in Additive Manufacturing, 2019. **4**(3): p. 245-267.
39. Sanandiya, N.D., et al., *Circular manufacturing of chitinous bio-composites via bioconversion of urban refuse.* Scientific Reports, 2020. **10**(1): p. 4632.
40. Byrne, G., et al., *Biologicalisation: Biological transformation in manufacturing.* CIRP Journal of Manufacturing Science and Technology, 2018. **21**: p. 1-32.

**Table of Contents**

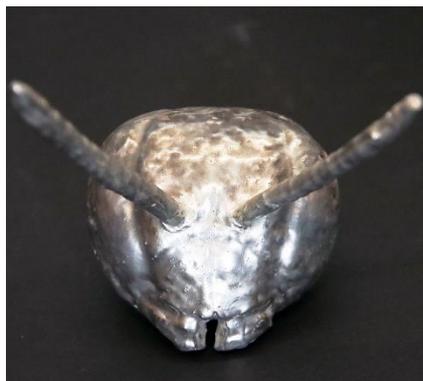
Biological systems have evolved to produce extraordinary materials with minimum energy and resources. By reproducing the use of metals in biological structures, a new paradigm in metalworking for the fabrication of ultra-low binder content metallic composites is demonstrated. The process occurs at room temperature and uses water as the only solvent, enabling the fabrication of standalone objects or their combination with biomaterials.

**Authors:** Ng Shiwei, Ng Guan Zhi Benjamin, Robert E. Simpson, Javier G. Fernandez

**Title:** A biological approach to metalworking based on chitinous colloids and composites

Supplementary Information

# A biological approach to metalworking based on chitinous colloids and composites

Ng Shiwei, Ng Guan Zhi Benjamin, Robert E. Simpson, & Javier G. Fernandez*

* To whom correspondence should be addressed. E-mail: javier.fernandez@sutd.edu.sg

This material includes:

**Materials and Methods**

    **Preparation of metallic chitosan composite**
    **X-Ray Diffraction (XRD)**
    **Scanning Electron Microscopy (SEM) & Energy-Dispersive X-ray Spectroscopy (EDS)**
    **X-ray microtomography**
    **Flexural strength measurement**
    **Coating samples and loading measurement**
    **Differential Scanning Calorimetry**
    **Electrical Resistance as a function of temperature measurement**
    **Additive manufacturing of 3D objects**

**Tables**

    **Table S1** – Physical properties and composition change with and without heat treatment of Chitosan-Tin composites
    **Table S2** – Resistance across the samples for different metals (See also Figure S3)

**Figures**

    **Figure S1** – Cold casting and thermal stability of metal pieces (Copper)
    **Figure S2** – Composite stability after water immersion
    **Figure S3** – Thermal and electrical tests
    **Figure S4** – Mercury intrusion plots
    **Figure S5** – Printing set-up

**Videos**

    **Video S1** – Electrical continuity test
    **Video S2** – Production of electrically conductive biological objects at ambient conditions.
    **Video S3** – Easy production of conductive tracks in chitometallic composites.
    **Video S4** – Bioprinting of a functional E27 bulb socket

# Materials and Methods

**Preparation of metallic chitosan composite**

To produce a material that can be economically scaled, accessible and relevant outside of laboratory environment, the materials sourced were in their unideal but most common form; And experiments were performed with a focus on its practical application. As the second most ubiquitous organic material in the world and a part of every ecosystem, potential sources of chitin are numerous; especially after the demonstration of its extraction from Black Soldier Fly which facilitates bioconversion of organic waste in urban environments. In our experiments, we used deacetylated chitin from the abundant processing waste of shrimp in South East Asia. Commercially available chitosan (medium molecular weight; 75–85% deacetylated) was sourced from i-CHESS chemicals Pvt Ltd., India and dispersed in 1% (v/v) acetic acid to form 3% (w/w) chitosan solution. The initially viscid solution was submerged in a 50℃-water bath for 72 hours to fully disperse chitosan.

On the other hand, as received metal powders produced by solid-state reduction, with suboptimal purity and variable geometry were used. Metallic powders were sourced from Chengdu Best New Materials Co., Ltd and used as received. Tin powder (~200 mesh, 99.6% Sn), Copper powder (~250 mesh, 99.5% Cu) and Stainless steel 304 powder (~300 mesh, C≤0.07 ,Si≤1.0 ,Mn≤2.0 ,Cr 18.0-19.0,Ni 8.0-11.0, S≤0.03, P≤0.035) were mixed in varying (w/w) ratios from 1: 100 to 1:250 for different tests and demonstrated applications.

**X-Ray Diffraction (XRD)**

XRD measurement was conducted using Bruker Eco D8 Advance diffractometer at CuK$\alpha$ radiation (wavelength = 1.542Å) at a voltage of 40 kV and 25mA. Scans between a 2$\theta$ range of 5° and 55° were performed with at step size of 0.02° and 5°.min$^{-1}$. Presented data were results of post processing on DIFFRAC.EVA software that includes stripping K$\alpha$2 contributions, smoothening and removing background.

**Scanning Electron Microscopy (SEM) & Energy-Dispersive X-ray Spectroscopy (EDS)**

SEM observations of surface morphology were made using a Field Emission Scanning Electron Microscope (FESEM, JEOL, JSM-7600F) at 15kV accelerating voltage. Samples were affixed on a carbon tape onto an aluminum stub and sputter coated with gold for 30s (5-10nm thickness). The SEM images were cropped, levels adjusted and colorized manually using MountainsSEM® colorization tool.

SEM-EDS observations of pre- and post-heat-treated samples were produced with an acquisition time of 12 minutes in the same conditions to detect elemental changes and layered onto SEM images.

**X-ray microtomography**

To understand better the internal structure of composite, 3D imaging was performed using Bruker SkyScan 1272. Scans were conducted with a source voltage of 100kV and source current of 100uA.

**Flexural strength measurement**

Metallic chitosan composite of 1:50, 1:00, 1:150, 1:200 and 1:250 chitosan to metal powder (w/w) ratio were prepared, casted into silicone molds (96mm (length) by 13mm (width) by 4mm (thickness)) and dried at room ambient conditions. Specimen span length (1:20) were determined in accordance with ASTM D7264M. Samples were subjected to a loading rate of 2.67mm/min in accordance to ASTM D790 procedure A. All experiments were performed in Singapore (15 m A.M.S.L.) at laboratory conditions: 20 ° C, 68% R.H. Experimental data were expressed as the mean ± standard deviation, with at least triplicate measurements.

**Coating samples and loading measurement**

Dried FLAMs samples (1:8 (w/w) chitosan to cellulose ratio) were cut into short bars of 7 mm by 7 mm by 5.5 mm as penetrating samples while flat sheets of 110 mm by 41 mm by 2mm were prepared as penetrating target. Penetrating samples were sharpened with a off-shelf pencil sharpener to obtain a sharp tip. For coated samples, the sharpen tip was then dipped into a chitosan-tin mixture of 1:250 (w/w) ratio and allowed to drip dry. Similarly, casted "Merlion" and honey bee was coated. After drying the coated surfaces were polished.

The penetration target was placed on a flexural jig with supports at 3 mm length span. The coated and uncoated sharpen tips were secured between pneumatic actuated grips and pushed into the target FLAM at a rate of 4 mm/min. Load and displacement data were recorded.

**Differential Scanning Calorimetry**

Heat flow measurements between 30°C to 500°C were conducted using DSC Q20 (TA Instruments) at a ramp rate of 10°C/min. Upon reaching 500°C, an isothermal state was maintained for 2 mins and subsequently ramped down at the same rate.

**Electrical Resistance as a function of temperature measurement**

The electrical resistance of 1 cm by 1 cm by 3 mm chitometallic samples was measured by heating the samples in air using a Torrey Pines computer-controlled hotplate, a Keithley 2410 source-measure unit (SMU), and a four-point probe (4PP) with spring-loaded tungsten tips. The tungsten probe tips were spaced 5 mm apart along a line. Since the hotplate had an aluminum surface, a glass slide was used to electrically isolate the sample from the hotplate. The current, $I$, was delivered to the sample through the outer two probes of the 4PP, whilst the inner two probes were used to measure the potential difference, $\Delta V$, due to the current flowing through the sample. This measurement arrangement avoids contact resistance issues. The Keithley 2410 was programmed to maintain a constant voltage of 2.1 V potential difference across the inner two probes by adjusting the current delivered through the outer two probes. The sheet resistance was calculated as $R_s = \frac{\pi}{ln(2)} \frac{\Delta V}{I}$. The hotplate was programmed to increase in temperature at a rate of 5 $^0$C/min to a maximum temperature of 350 $^o$C. The current through the outer probes and the voltage drop across the inner probes was recorded every second whilst the sample was heated and whilst it cooled naturally. When the samples underwent the chitometallic to metal phase transition, there was a radical reduction in the electrical sheet resistance, and therefore the Keithley 2410 SMU compensated by substantially increasing the current delivered to the outer probes such that the voltage drop across the inner probes was maintained at 2.1V. Thus, the resistance is measured to decrease at this transition temperature.

**Additive manufacturing of 3D objects**

A solid 3D model of a light bulb socket was designed in Autodesk Fusion 360, then hollowed out and sliced with Ultimaker Cura using the Spiralize Outer Contour Mode to produce gcode for the printer. Printing was achieved using a modified CNC router with a cartesian motion system and a syringe-loaded pneumatic paste extruder. A chitosan-tin mixture of 1:250 ratio was loaded into a 150 ml syringe and extruded at a pneumatic pressure of 8 to 10 psi. The material was extruded on an aluminum foil coated hot plate which was heated at 60°C and progressively increased to 100°C during printing. This facilitated drying and improved the buildability of material layer by layer. End connector wires were placed mid-print to make the eventual socket directly functional. Once the printing was complete, the socket was dried for 15 minutes at 100°C to expel residual moisture. The hot plate was then heated to 320°C to thermally treat the socket, which reduced its internal resistance. The socket was then cooled to ambient temperature and connected to 12V and ground. A 12VDC 7W E27 bulb was then mounted.

# Tables

| Properties | | Without heat treatment | With heat treatment |
|---|---|---|---|
| Bulk Density (g/cm$^3$) | | 4.4569 | 5.2766 |
| Apparent (skeletal) Density (g/cm$^3$) | | 6.7843 | 6.5623 |
| Interstitial Porosity (%) | | 34.3052 | 25.95 |
| C,H,N Elemental Analysis | Carbon (%) | 0.25 | 0.06 |
| | Hydrogen (%) | 0.04 | Not detected |
| | Nitrogen (%) | 0.06 | Not detected |

**Table S1| Physical properties and composition change with and without heat treatment of 1:150 Chitosan-Tin composite.** Interstitial porosity reduced after heat treatment, confirming further internal compaction of composite as chitosan undergoes thermal oxidative degradation. This is supported by a reduction in carbon content and non-detectable hydrogen and nitrogen content after heat treatment. As overall volume reduces and internal packing increases, bulk density increases. Without accounting for interstitial space between particles, apparent density remains similar before and after heat treatment with a small decrease that might correspond to the closure of some pores.

| Material | Before heat treatment ($\Omega$) | After heat treatment ($\Omega$) |
|---|---|---|
| Chitosan (film) | 1.29 ± 0.22 ×10$^8$ | 8.67 ± 2.66 ×10$^8$ |
| Tin | 4.27 ± 0.51 ×10$^8$ | 1.14 ± 0.92 ×10$^{-2}$ |
| Copper | 2.21 ± 1.51 ×10$^8$ | 1.63 ± 0.13 ×10$^2$ |
| Steel | 2.01 ± 0.68 ×10$^8$ | 6.23 ± 3.33 ×10$^6$ |

**Table S2| Electrical resistance of the samples before and after heat treatment.** Resistance of the different chitometalllic porous composites before and after heat treatment, measured in the 50-100 °C range. For a detailed and continuous description of the transition refer to Figure S3. Errors are the standard deviation.

# Figures

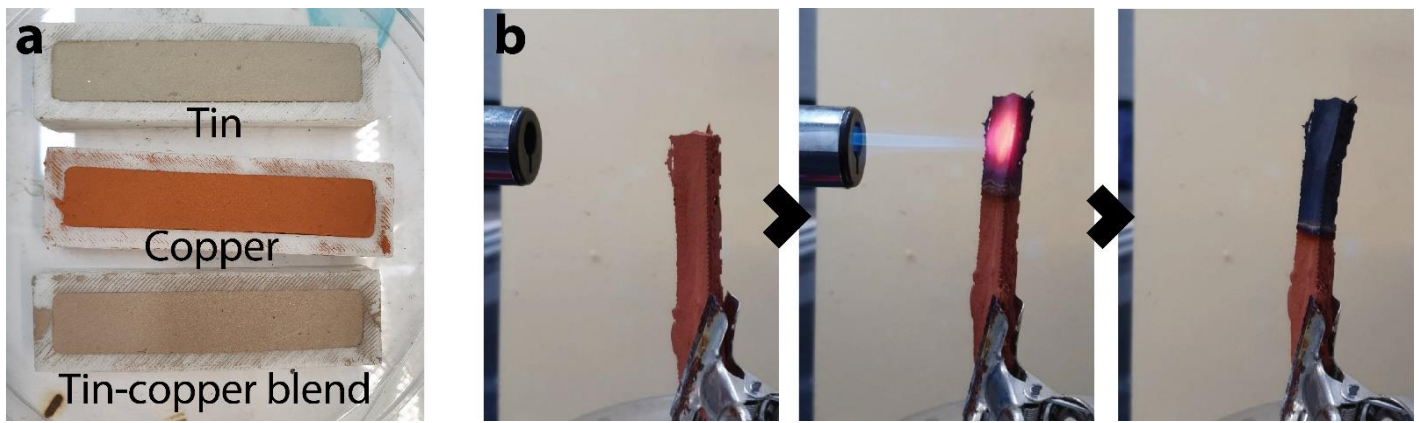

**Figure S1| Cold casting and flame test. a,** Examples of chitometallic slabs produced at room conditions by casting colloids in silicone molds. **b,** The sequence presents the effect of a butane flame (~1500 °C) on a chitosan-copper slab produced at ambient conditions. Even at those temperatures, well above those for the carbonization/vaporization of most organic components, the chitometallic aggregates remain solid and preserve their geometry.

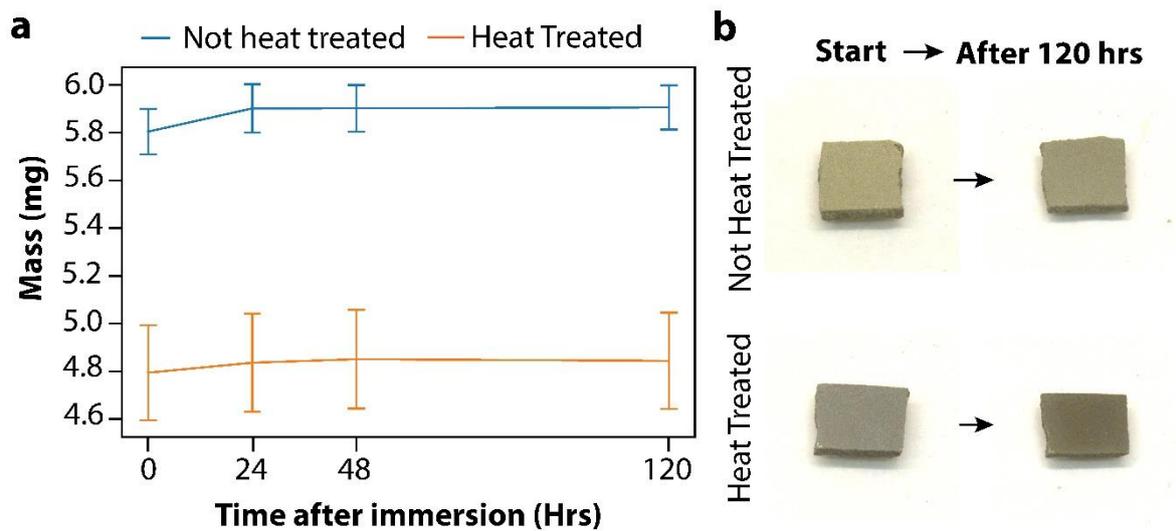

**Figure S2| Composite stability after water immersion. a.** Rather than dissolving, the composite gained mass right after immersion and plateaued after 24 hours, with seemingly lesser water uptake by the heat-treated sample. **b.** Composite remains stable even up to 120 hours of immersion, without observable loss of material or swell in dimensions.

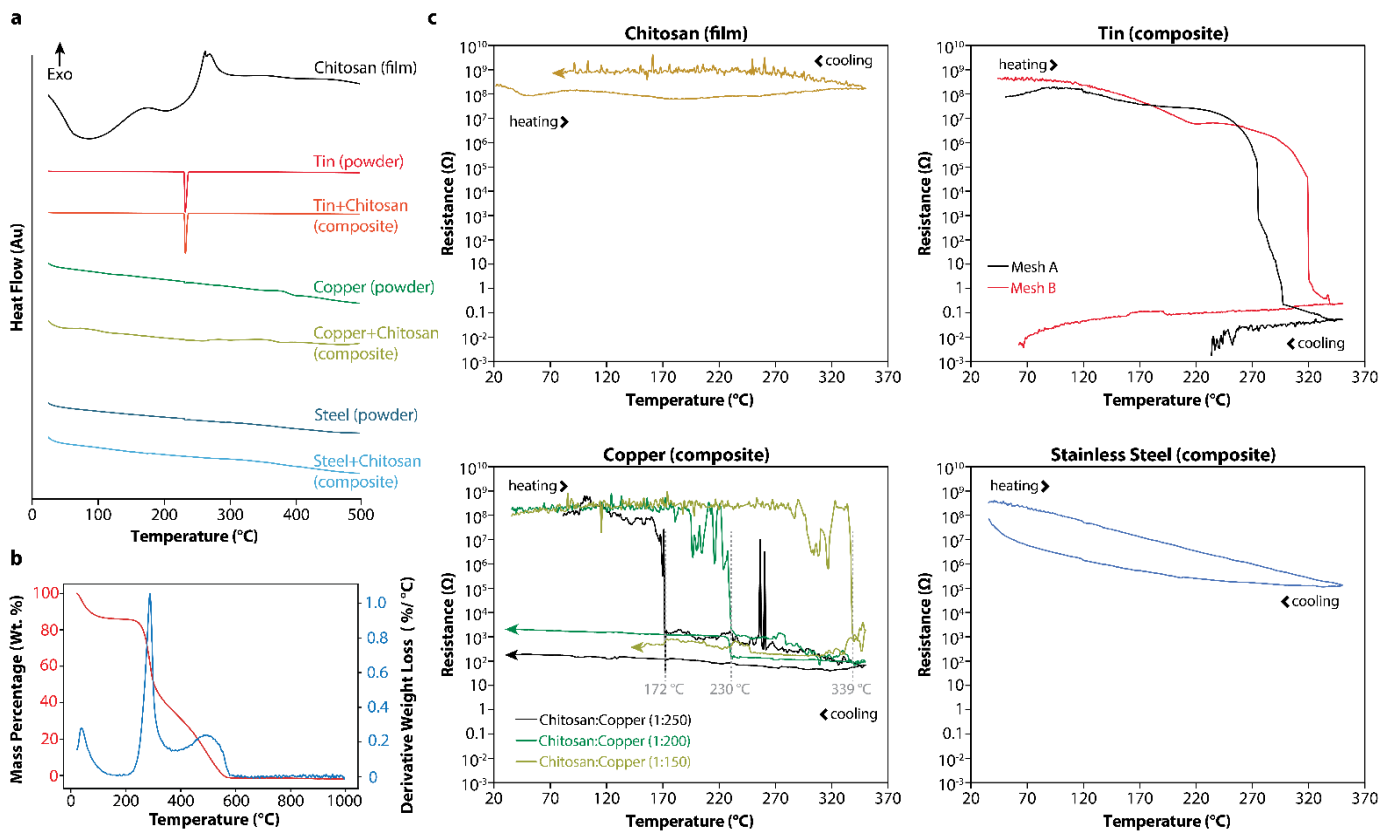

**Figure S3| Thermal and electrical tests. a,** Differential scanning calorimetry (DSC). Of all the metals used, only tin has an endotherm peak at its melting temperature. Due to the extremely low amounts of chitosan in the composites, there are no observable differences between the heat flow of the metal alone and the chitometallic composites. **b,** Thermogravimetric analysis (TGA) of chitosan films in air shows three stages of weight loss. The first stage below 100°C relates to the initial dehydration of film, followed by drastic weight loss in the range of 280°C to 300°C (degradation of acetylated/ deacetylated units, depolymerization and dehydration of saccharide rings) and finally the complete thermal oxidative degradation before 600°C **c,** Electrical Resistance and thermal treatments. The data represents the electrical properties of the different samples while heated to 350°C on a hot plate (open atmosphere) and their cooling down to room temperature. The arrows in the graph represent the direction of the cycle (i.e., from left to right is the heating process, and from right to left the cooling back to ambient temperature). Thermally treating chitosan films alone gives rise to a more resistant material, suggesting that carbonization of the organic binder does not play a relevant role in the process. Tin and copper have a permanent drop in resistance of about 10 and 6 orders of magnitude, respectively. In contrast, Stainless Steel shows an order of magnitude change. It is worth noting that the size of stainless steel particles is several times larger than those of the other metals (see Fig. 1c). The transition occurs sharply at temperatures unrelated to the metal used (see panel a). For the same metal, changing the ratio of chitosan to metal (See graph for copper) or the size of the metallic particles (See graph for tin) results in shifts in the electrical transition temperature. All the data suggests that the process in which the chitometallic composites acquire the electrical properties of a metallic foam is independent of the metal used and is driven by a physical process. Because of the absence of external pressure and reliance on internal forces, the process could be described, for the lack of a better definition, as a "self-percolation" or "self-compaction" process.

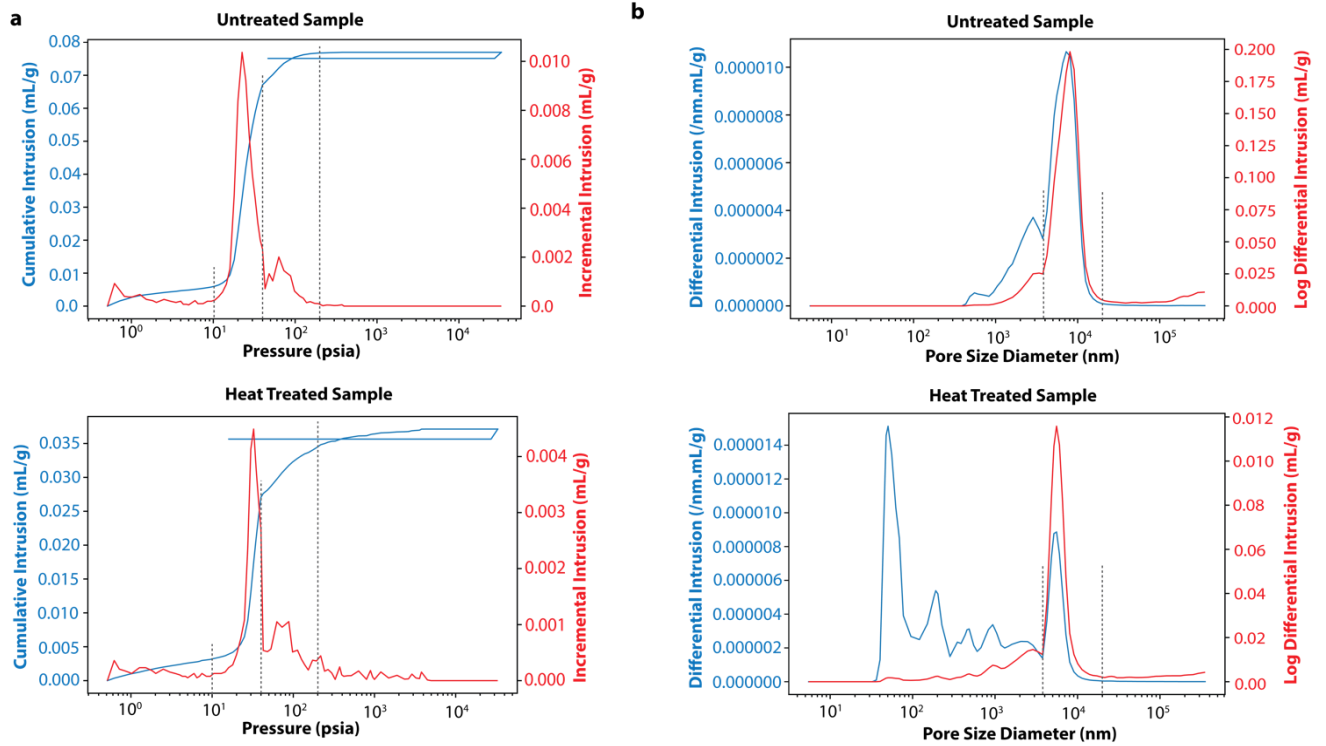

**Figure S4| Mercury intrusion plots. a.** Between the cumulative intrusion and incremental intrusion curve of untreated and heat-treated samples, it can be inferred that macropores persist after heat treatment. However, smaller pores are present in heat-treated samples as seen from a lower fraction of cumulative intrusion at 40psia and continuing mercury intrusion beyond 200 psia. **b.** The differential intrusion curve (1st derivative of the cumulative intrusion) plotted against pore size diameter illustrates this change in the distribution of pore sizes with a multimodal distribution after heat treatment.

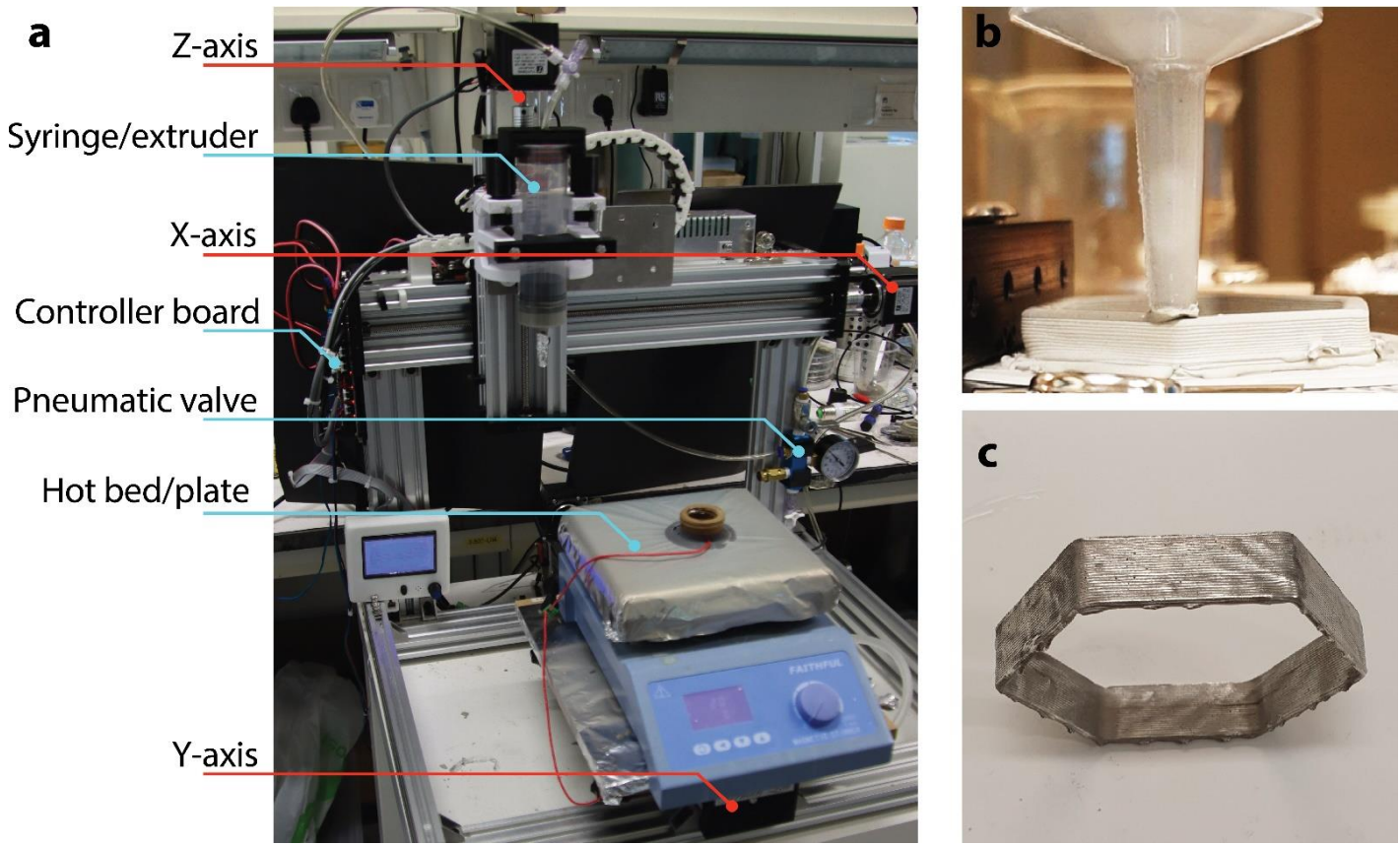

**Figure S5| Printing set-up. a,** The printing set-up consists of a pneumatically actuated syringe dispensing the chitometallic colloid and assembled on a cartesian positioner, printing on a hot plate. The hot plate is included to perform the printing and thermal treatment in the same set-up. However, the hot plate could be removed since printing and treatment are two disjunct processes (the latter optional). In a simplified set-up, the printing could be done on a flat surface, and the resulting sample thermally treated somewhere else. The lack of temperature during printing also enables low requirement parts, such as a regular polypropylene syringe for dispensing. The cartesian positioner is actuated by three stepper motors, one at the base controlling a horizontal axis, one at the syringe carrier controlling the second horizontal axis, and one controlling the vertical position of the syringe. **b,** Example of the process of printing a piece at room temperature. **c,** The printed piece of the previous panel after polishing. No thermal treatment has been performed in this case. For an example of thermal treatment, refer to Supplementary Video 4.

# Videos

**Video S1 – Electrical continuity test.** Testing, using a multimeter, the electrical conductivity of several chitometallic objects, produced by different methods and with and without different surface and bulk treatments.

**Video S2 – Production of electrically conductive biological objects at ambient conditions.** Metallic functionalization of a replica of a bee head and a Merlion made out of cellulose (Fungal-like adhesive materials) using a tin-based chitometallic colloidal suspension. The process, happening at ambient temperature and pressure, covers the path from the digital model to the functional biometallic object.

**Video S3 – Easy production of conductive tracks in chitometallic composites.** Demonstration of localized surface continuity on a chitosan-tin composite produced with a blunt tool.

**Video S4 – Bioprinting of a functional E26 bulb socket.** The printing process of a bulb socket at ambient conditions and its thermal treatment to achieve bulk electrical properties. The video covers, without cuts, the whole process from printing the first layer to using the socket to screw and light a commercial bulb.